\documentstyle[multicol,epsf,aps,amsmath,amssymb,amsbsy]{revtex}

\begin{document}
\title{Effects of Non-local Stress on the Determination of Shear
  Banding Flow} \author{C.-Y. David Lu$^{a}$ , Peter D. Olmsted$^b$,
  and R. C. Ball$^{c}$} \address{$^a$Departments of Physics and
  Chemistry \& Center of Complex
  Systems, National Central University, Chung-Li, 320, Taiwan; \\
  $^b$Department of Physics and Astronomy \& Polymer IRC,
  University of Leeds, Leeds, LS2 9JT, UK;  \\
  $^c$Department of Physics, University of Warwick, Coventry, CV4 7AL,
  UK} \maketitle

\begin{abstract} 
  We analyze the steady planar shear flow of the modified
  Johnson-Segalman model, which has an added non-local term. We find
  that the new term allows for unambiguous selection of the stress at
  which two ``phases'' coexist, in contrast to the original model.
  For general differential constitutive models we show the singular
  nature of stress selection in terms of a saddle connection between
  fixed points in the equivalent dynamical system.  The result means
  that stress selection is unique under most conditions for space
  non-local models.  Finally, illustrated by simple models, we show
  that stress selection generally depends on the form of the non-local
  terms (weak universality).
\end{abstract}
\vspace{-12mm}
\begin{multicols}{2}
  \textsl{Introduction}---Oldroyd's proposal \cite{Oldroyd} that any
  sensible rheological constitutive equation for a general fluid
  should obey the ``admissible conditions'' has had a great influence
  on the rheology community. In his ``admissible conditions'', apart
  from requiring covariance, Oldroyd further imposed the ``principle
  of frame indifference'' and ``locality''.  de Gennes has shown that
  the former is no longer true if inertia effects are important
  \cite{deGennes}. Below we discuss how one must extend locality to
  model shear banding flow, which appears in some surfactant
  \cite{Rehage,Berret,Schmitt94,Grand} and polymer \cite{Callaghan2}
  solutions.
  
Shear banding has long been mooted by the polymer rheology community
  in connection with the spurt effect in the processing of linear
  polymer melts (see, \emph{e.g.} Ref.\cite{Review} for a review). It
  is, however, in surfactant worm-like micelle solutions that the
  phenomena has been firmly established
  \cite{Rehage,Schmitt94,Berret,Grand}, particularly convincingly
  through magnetic imaging experiments \cite{Callaghan}.  For
  entangled polymers or surfactant micelles, as the shear rate
  increases the polymers/micelles gradually align and the fluid shear
  thins.  According to reptation theory \cite{Doi,Mike1}, the fluid
  shear thins so heavily that a maximum appears in the shear
  stress-shear rate relation at a shear rate of approximately the
  inverse reptation time.  At high enough shear rate, the shear stress
  is presumed to increase again, resumably due to local ({\em e.g.}
  solvent) dissipation mechanisms \cite{Neil2}. The steady shear rate
  curve is qualitatively like Fig.~1.  For intermediate shear rates
  where the slope is negative, homogeneous flow is mechanically
  unstable.  It is found in real systems that, when the controlled
  shear rate increases up to the order of the inverse reptation time,
  the fluid becomes a composite flow profile (shear bands), where two
  or more bands, with alternating low and high shear rates, coexist
  with a common shear stress.

  According to Fig. 1, there is a range of shear stress
  $[\sigma_{min},\sigma_{max}]$ within which the constitutive equation
  is multi-valued (three shear rates for a given stress), so that
  shear bands seem possible for any stress within
  $[\sigma_{min},\sigma_{max}]$. However, real systems select a well
  defined stress \cite{Berret,Grand}.  We therefore need a mathematical
  condition, a selection criterion, which determines the coexisting
  stress. There are scattered results showing generically that steady
  state solutions behave very differently depending on whether or not
  the constitutive equations are local in space \cite{foot1}. For
  local models, the final steady stress depends on the flow history
  \cite{History}.  Therefore an additional selection criterion must be
  imposed to select stress, be it variational \cite{Porte,Ball} or not
  \cite{Mike3,Lequeux}. On the other hand, steady state analysis of
  several non-local models, either analytically \cite{Neil1,Pearson}
  or numerically \cite{Olmsted1,Olmsted2}, yields a well-defined
  selected stress. It is interesting to know whether  non-local
  effects generally lead to a stress selection criterion.

{\narrowtext
{\begin{figure}[!tbh ]
\displaywidth\columnwidth
\epsfxsize=3.5truein
\centerline{\epsfbox[140 310 450 510]{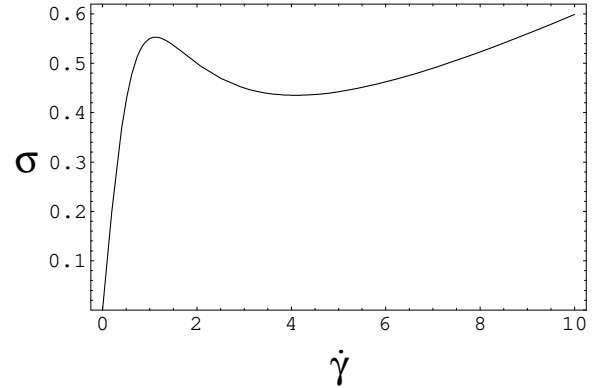}}
\caption{Local constitutive relation $\sigma(\dot \gamma)$ 
  for the Johnson-Segalman (JS) model (Eqs.~\ref{JS1}-\ref{JS3}) with
  $e=0.05$.}
\end{figure}}
}

In this paper, we first present numerical results for the
modified Johnson-Segalman (JS) model, as a concrete example of
stress selection in a non-local model that has been well-studied
in local form by many workers \cite{Malkus,Renardy,Ball}. Although
perhaps not molecularly faithful to any fluid, the JS model is a
covariant model which posesses a non-monotonic flow curve, 
providing the necessary ingredient for a qualitative study of shear 
banding. We then present the main result which shows the general 
link between non-locality and stress selection.  Finally we use simple 
examples to illustrate the important fact that stress selection depends 
on the details of the interface structure, \emph{i.e.}  it has weak universality.

\textsl{Numerical Results for the Modified JS Model with Stress
  Diffusion}---This model is given by
\begin{eqnarray}
  \rho (\partial_{t}+{\bf v}\cdot \boldsymbol{\nabla}){\bf v} &=&
  \boldsymbol{\nabla}\cdot 
  \boldsymbol{\sigma } 
  \label{1} \\
  \boldsymbol{\sigma} &=&-p\,{\bf I} + 2 \eta \boldsymbol{\kappa } +
  \boldsymbol{\Sigma}  \label{2} \\
  \overset{\blacklozenge}{\boldsymbol{\Sigma}}
  &=&2 G\boldsymbol{\kappa }-\boldsymbol{\Sigma }/\tau
  +D\nabla ^{2}\boldsymbol{\Sigma }  
  \label{3}
\end{eqnarray}
where the strain rate is $\boldsymbol{\kappa} \equiv
(\boldsymbol{\nabla}{\bf v} + \boldsymbol{\nabla}{\bf v}^{T})/2$.  The
first equation is the momentum equation, with $\rho, {\bf v},
\boldsymbol{\sigma}$ the fluid density, velocity, and stress tensor,
respectively.  The stress comprises the pressure $p$ ({\bf I} denotes
the identity matrix), the elastic stress $\boldsymbol{\Sigma }$, and a
Newtonian viscous stress with viscosity $\eta$. We assume an
incompressible fluid, $\boldsymbol{\nabla}\cdot{\bf v}=0$, enforced by
the pressure. The non-Newtonian stress $\boldsymbol{\Sigma }$ is
governed by Eq.~(\ref{3}), where the stress is induced by strain rate
$\boldsymbol{\kappa}$ with strength described by a modulus $G$, and $\tau$ is the
stress relaxation time. The Gordon-Schowalter convected time
derivative is
\begin{eqnarray}
  \overset{\blacklozenge}{\boldsymbol{\Sigma}}&\equiv&
  \left(\partial_{t} + {\bf v}\cdot\boldsymbol{\nabla}\right)
  \boldsymbol{\Sigma}-\boldsymbol{\nabla}{\bf v}^{T}\cdot
  \boldsymbol{\Sigma }- \boldsymbol{\Sigma} 
  \cdot \boldsymbol{\nabla}{\bf v} \\
&& + (1-a) \left(\boldsymbol{\Sigma }\cdot
  \boldsymbol{\kappa }+\boldsymbol{\kappa }\cdot \boldsymbol{\Sigma
  }\right), \nonumber
\end{eqnarray}
and describes the ``slip'' between the polymer and fluid extension,
with $a \in \lbrack -1,1]$.

Eqs.~(\ref{1}-\ref{3}) differ from the standard JS model \cite{Larson}
due to the additional stress diffusion term $D\nabla
^{2}\boldsymbol{\Sigma }$ in Eq.~(\ref{3}).  El-Kareh and Leal
\cite{Leal} have analyzed the microscopic dumbbell model (\emph{i.e.}
the case $a=1$) and derived this term, representing the diffusion of
the stress carrying element. Note the important difference that the
standard JS model is a local model, whereas Eq.~(\ref{3}) is not.

Given a shear stress value $\sigma_{xy}$, we look for the steady
planar shear flow numerically. For ${\bf v}=v (y,t) \hat {{\bf e}}_x$,
the modified JS model reduces to \cite{Ball}
\begin{eqnarray}
\sigma&=& S + e \dot{\gamma}  \label{JS1} \\
\tau \partial_{t} S &=& \hat D \partial^2_y S - S -\dot{\gamma} N+ 
\dot{\gamma}  \label{JS2} \\
\tau \partial_{t} N &=& \hat D \partial^2_y N - N +\dot{\gamma} S 
\label{JS3}
\end{eqnarray}
The rescaled
variables are defined as $ \sigma=\sqrt{1-a^2} \ \sigma_{xy}/G$,
$S=\sqrt{1-a^2} \ \Sigma_{xy}/G$,
$N=[(1-a)\Sigma_{xx}-(1+a)\Sigma_{yy}]/2G$, $\hat D=\tau D$,
$\dot{\gamma}= 2\sqrt{1-a^2} \ \tau \kappa_{xy}$, and $e=\eta/G \tau$.
Because the steady planar shear flow solution of Eq.~(\ref{1}) has
zero Reynolds number (and most macromolecular dynamics are in this
limit) we can use the low Reynolds number limit ($\rho\rightarrow 0$),
Eq.~(\ref{JS1}), to obtain the same steady solution.
Eqs.~(\ref{JS2}-\ref{JS3}) are the non-trivial dynamics of
$\boldsymbol{\Sigma}$ (other components are uncoupled).  
Note that there is only one non-dimensional parameter $e$ (the
viscosity ratio) in the steady state problem.

Given $\sigma$, we integrate Eqs.~(\ref{JS1}-\ref{JS3}) over time $t$
to see whether a steady state banding solution can be reached.  During
the integration, the values of $S$ and $N$ at the two ends $y=\pm
y_{0}$ are fixed to those of the high and low strain rate branches
respectively. At $t =0$, the functions $S(y) $ and $N(y)$ are chosen
arbitrarily as $C_{1}\tanh y+C_{2}$, with the constants $C_{1}$, $C_{2}$ fixed
by the boundary conditions.  The value $y_0$ is chosen to be
sufficiently large that its exact value is irrelevant. We find that a
steady (elementary) shear band solution can be found only at a
specific shear stress value, $\sigma=\sigma_{0}$.  If the stress value
is too large or too small, one of the phases shrinks completely.  The
values $\sigma_{0}$ as a function of the model parameter $e$ are shown
in Fig.~2. Note that in an infinite system $\sigma_{0}$ does not
depend on the diffusivity $D$, which just sets the interface width.
However, anisotropic diffusion (a fourth rank tensor
$\boldsymbol{D}$ in Eq.~\ref{3}), may lead to diffusivity dependent selection.

{\narrowtext
\begin{figure}[!tbh]
\epsfxsize=3.2truein
\centerline{\epsfbox[-30 475 420 765]{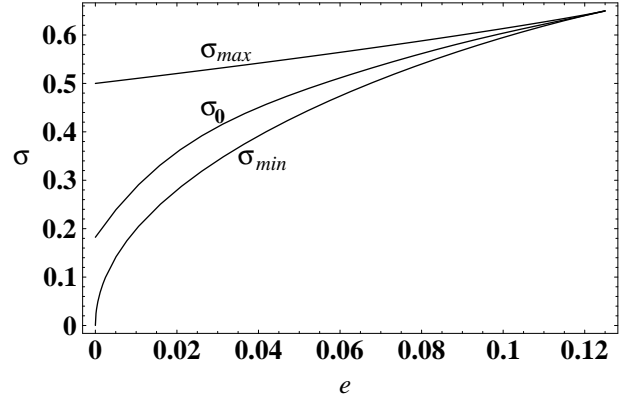}}
\caption{The selected stress $\protect\sigma_{0}$
  of the modified Johnson-Segalman model as a function of the
  viscosity ratio $e$. The limits $\protect\sigma_{min}(e)$
  and $\protect\sigma_{max}(e)$ are also shown. }
\label{Fig2}
\end{figure}
}

\textsl{Nontransverse Saddle Connection}---We have shown above that a
non-local model can select the stress sharply.  Now we show that, in
general, a non-local model in planar shear flow selects stress sharply
(could be uniquely), provided that the model is a differential
equation and satisfies rotational and Galilean invariance.

\begin{enumerate}
\item To begin, we observe that the steady state equations, like
  Eqs.~(\ref{JS1}-\ref {JS3}) without time derivatives, comprise a set
  of {\it ordinary} differential equations (ODE) for the
  independent dynamical variables $\{\psi_i\}$ ($S$ and $N$ for the JS
  model), rather than partial differential equations (PDE), since only
  differentiation in the velocity gradient direction, say
  ${\bf\hat{y}}$, is present in planar shear flow.  The phase space
  for the equivalent first order ODE is
  $\left[\left\{\psi_i\right\},\left\{\partial_y\psi_i\right\},
    \left\{\partial^2_y\psi_i\right\},\ldots\right]$.  Higher order
  gradients in the original PDE would entail a ODE phase space with
  large dimension.

\item To each homogeneous steady flow (a phase) there
  corresponds a fixed point
  ($\left[\left\{\psi^{\ast}_i\right\},0,0,\ldots\right]$) 
in the ODE phase space since, by definition, homogeneous flow means the
  variables do not change with $y$ \cite{foot3}.
  
\item  If the interface width is small compared to the other length
  scales in the problem, \emph{e.~g.} the distance between two
  neighbouring interfaces, it is sufficient to consider an elementary
  shear banding solution, which describes, in a
  planar geometry, a composite flow with a smooth interface separating
  a single region of material in the high shear rate phase from a
  single region of material in the low shear rate phase.
  Mathematically, the elementary shear band is a solution of the ODE
  which asymptotically connects the high and low shear rate phase
  fixed points between $y=\pm \infty$. Since the fixed points are both saddle points (see
  below) and distinct, the elementary shear band solution is a \emph{heteroclinic saddle
    connection} \cite{saddle}.
  
\item Let the attractor and repellor basins \cite{saddle} of one fixed point, say $A$, have
  dimensions $a(A)$ and $r(A)$ respectively. Another fixed point $B$
  has similarly defined $a(B)$ and $r(B)$. Note that if the phase
  space has dimension $d$ (for the modified JS model $d=4$), then
  $a(A)+r(A) \leq d$ and $a(B)+r(B) \leq d$.  A saddle connection
  joins two fixed points, so that it is an intersection of the
  repellor of one fixed point, say $A$, and the attractor of the other
  fixed point $B$. We denote its dimension $s(A,B)$ \cite{foot4}.  In
  the ODE phase space, the intersection is at least one dimension, so
  that $s(A,B) \geq 1$.  There is a trivial inequality
  $r(A)+a(B)-s(A,B) \leq d$.  If equality holds the saddle connection
  is called {\em transverse\/}: it is formed by a robust intersection
  of two manifolds, and is structurally stable against small
  perturbations of the ODE parameters (in the modified JS model,
  $\sigma$ and $e$).  The case $r(A)+a(B)-s(A,B) < d$ is called {\em
    non-transverse\/}, and the corresponding saddle connection is
  structurally unstable. It is important to recognise that if the
  elementary shear band solution is a non-transverse saddle
  connection, a small change of the shear stress (a parameter in the
  ODE) will remove the saddle connection, which gives a very sharp
  stress selection. In another words, given an existing shear band
  solution, a stress perturbation is singular if it is of the
  non-transverse type saddle connection. On the other hand, if there
  is a transverse saddle connection shear band solution, one can
  perturb the stress to obtain a (slightly) different shear band
  solution.  Therefore, {\em there is no stress selection for a transverse
  saddle connection}. (Stress perturbation becomes a regular
  perturbation.)
  
\item We now prove that: if the model has rotational and Galilean
  invariance, a shear band solution in planar shear flow must be
  non-transverse.
 Let us momentarily assume that there is a transverse saddle
  connection approaching, say, fixed point $A$ at $y = -L$ to fixed point
  $B$ at $y =L$. Galilean invariance allows us to choose an inertial frame
  in which the flow velocity $v_x(y=\pm L)=\pm V$.  Since the bulk are
  rotationally invariant and the boundary conditions are symmetric
  under rotation by 180 degrees, one can rotate the solution around
  the vorticity ($z$) axis by 180 degrees to obtain a different shear
  band solution which obeys the same equation and the same boundary
  condition ({\em e.g.} ${v_x (\pm L)=\pm V}$). This solution approaches
  fixed point $B$ at $y = -L$ to fixed point $A$ at $y =L$.  Fixed
  points $A$ and $B$ have both attractor and repellor directions, and
  are thus saddle points. Now take the (thermodynamic) limit $L \rightarrow \infty$ so that the above two solutions become saddle connections.
  The two saddle connections are related by
  the symmetry transformations, so are of the same type (both
  transverse).  A contradiction now appears, since ``there are not
  enough dimensions to put two transverse saddle connections in the
  phase space''.  Formally, the transverse condition leads to
  $r(A)+a(B)=d+s(A,B) > d$ and $r(B)+a(A)=d+s(B,A) > d$, therefore
  $a(A)+r(A) + a(B)+r(B) > 2d$.  However, from $a(A)+r(A) \leq d$ and
  $a(B)+r(B) \leq d$ we have $a(A)+r(A) + a(B)+r(B) \leq 2d$. So the
  shear band cannot be a transverse saddle connection.
\end{enumerate}

\textsl{Weak Universality}---One important aspect of stress selection
can be illustrated by a simple model inspired by the macromolecular
character of many systems exhibiting non-linear rheology. Let the
shear stress $\sigma$ be
\begin{equation}
  \label{eq:1}
  \sigma = F(\dot{\gamma}) + \eta\,\dot{\gamma}_{loc},
\end{equation}
where the first, non-linear, part associated with macromolecules is
sensitive to a locally averaged shear rate $\dot{\gamma}$ as opposed
to the second, Newtonian, part attributed to solvent, which is
sensitive to the true local shear rate $\dot{\gamma}_{loc}$. We
approximate the local averaging from $\dot{\gamma}_{loc}$ to
$\dot{\gamma}$ by a gradient expansion,
$  \dot{\gamma} = \left(1 - R^2(\dot{\gamma})\nabla^2 +
  \ldots\right)^{-1} \dot{\gamma}_{loc}$, 
and hence
\begin{equation}
  \label{eq:3}
  \dot{\gamma}_{loc} \simeq \left(1 - R^2\left(\dot{\gamma}\right)
  \nabla_y^2\right)  \dot{\gamma}
\end{equation}
We anticipate sensitivity of the non-local scale $R(\dot{\gamma})$ to
$\dot{\gamma}$ through distortion of the macromolecular shape.

In terms of the locally averaged shear rate $\dot{\gamma}$ we have 
\begin{eqnarray}
  \sigma = f(\dot{\gamma}) - \eta R^2(\dot{\gamma})\nabla_y^2\dot{\gamma}
  \label{eq:5}
\end{eqnarray}
where $f(\dot{\gamma})\equiv F(\dot{\gamma}) + \eta\,\dot{\gamma} $ is, as before, the steady (and constant) shear
rate flow curve as exemplified in Fig.1. Multiplying Eq.~(\ref{eq:5})
by  $R^{-2}(\dot{\gamma})\nabla_y\dot{\gamma}$ and
integrating $y$ across the shear band leads to 
\begin{eqnarray}
  {\cal D}(\sigma) \equiv
  \int^{\dot \gamma (\infty)}_{\dot \gamma (-\infty)} 
  \frac{f(\dot \gamma) -\sigma}{R^2(\dot \gamma)}d \dot \gamma
  =\left[ \frac{\eta}{2} (\nabla_y \dot \gamma (y))^2
  \right]_{y=-\infty}^{y=\infty}
  \label{4.2}
\end{eqnarray}
where $\dot \gamma (y=\pm \infty)$ depends on $\sigma$ through 
$f(\dot \gamma (y=\pm \infty))=\sigma$. 
Since $\nabla_y \dot \gamma (y=\pm \infty)=0$, an interfacial profile 
must satisfy ${\cal D}(\sigma)=0$, which is the condition to select the stress. 
 
According to Eq.~(\ref{4.2}), different functions $R(\dot \gamma)$
give different ${\cal D}(\sigma)$, and hence different selection
criteria ${\cal D}=0$! Therefore, two models with different $R(\dot
\gamma )$ but the same behaviour in homogeneous flow, $f(\dot
\gamma)$, can behave differently upon forming shear bands. The simple
case $R(\dot{\gamma})$ independent of $\dot{\gamma}$ corresponds to
the equal areas construction speculated upon by Ref.~\cite{Ball}, and
cannot be regarded as generic. Stress selection has weak universality,
implying that impurities or other effects which changes the
interfacial properties could in principle affect quantities like
$R(\dot \gamma)$ and hence alter the selected shear stress.

For equilibrium transitions, with coexistence equations analogous to
Eq.~(\ref{eq:5}), terms involving gradients would be exactly
integrable without an integrating factor ($1/R^2(\dot{\gamma})$ in
Eq.~\ref{4.2}) because they arise from a functional derivative of a
coarse-grained free energy. The resulting interface solvability
condition (\emph{i.e.} a Maxwell construction) is independent of the
detailed gradient terms, in contrast to the weak universality
discussed above.
\textsl{Discussion}---We close with a few comments.  First, we have
not proven existence of shear banding solutions.  There are physical
phenomena where coexistence between bulk states is prohibited because
it is unfavourable to form an interface, and robust hysteresis can be
expected. Gel swelling provides a beautiful example.

Secondly, the non-transversality condition of the saddle connection
is, strictly speaking, weaker than uniqueness. There are two possible
situations where uniqueness may fail.  The first situation happens
when, due to an accidental situation, the attractive basin of one
fixed point and the repelling basin of the other fixed point move so
as to maintain the saddle connection upon changing the shear stress
$\sigma$ as a control parameter.  The second situation happens
when more than one isolated stresses are selected, \emph{i.e.}
uniqueness fails globally.  These have negligible chance to be
realised in a model.  Should it happen, one may ask for a physical
argument for the degeneracy to be sure that it is not a mathematical
artfact. Generally, uniqueness of the stress selection can be expected for models with gradient terms

An interesting question is whether the stress selected by non-local
effects can be obtained by a variational principle.  A conventional
variational principle, like the one used for the equilibrium phase
transitions, relies on the volume contribution to a universal functional and
gives a criterion insensitive to the interfacial structure
\cite{foot2}, because the interface only contributes a vanishing
fraction if the total volume is assumed to be large.  The model
Eq.~(\ref{eq:5}), illustrates that the way non-local effects select
stress is different from a variational principle based on such a universal
functional. Therefore the obvious choices of either free energy or
entropy production cannot generally represent the selection criterion
posed by the non-local effects.

After this manuscript was submitted, Yuan published  a
similar modification to the JS model \cite{Yuan}.

\textsl{Acknowledgments}---We thank  A. Ajdari, M. Cates, 
B.~L. Hao, and O. Radulescu for fruitful conversations.  

\vspace{-5mm}

\end{multicols}

\begin{references}
\bibitem{foot1} From our point of view, 2D flow solver results which
  show a selected stress for a local model \cite{Neil1} should be
  examined further to see whether the algorithm implicitly introduced
  non-locality (numerical diffusion).
\bibitem{foot3} Note that although the velocity is not a constant, it
  cannot enter the equation of motion alone, due to Galilean
  invariance.
\bibitem{foot2} It is possible to have a functional whose bulk
  contribution is indifferent to the interface location, in which case
  the interface contribution plays the decisive role.
\bibitem{foot4} We assume $A$ and $B$ are the only singular points on
  the saddle connection, so that $s(A,B)$ does not change along the
  connection.
\end{references}
\end{document}